\documentclass[journal=mamobx,manuscript=article]{achemso}
\usepackage{titlesec}
\usepackage{fancybox}
\usepackage{mathrsfs}

\usepackage{setspace}
\usepackage{here}

\usepackage{comment}
\usepackage{enumerate}
\graphicspath{{./pdf/}}

\usepackage{amsmath}
\usepackage{cases}

\title{Nucleation theory of polymer crystallization\\ with conformation entropy}
\author{Hiroshi Yokota}
\email{hiroshi.yokota@riken.jp}
\affiliation{RIKEN Interdisciplinary Theoretical and Mathematical Sciences (iTHEMS),
Wako, Saitama 351-0198, Japan}
\author{Toshihiro Kawakatsu}
\affiliation{Department of Physcs, Tohoku University, Sendai, Japan}

\begin{document}
\maketitle
\section*{Abstract}
Based on classical nucleation theory, we propose a couple of theoretical models for the nucleation of polymer crystallization, i.e. one for a single chain system (Model S) and the other for a multi-chain system (Model M). In these models, we assume that the nucleus is composed of tails, loops and a cylindrical ordered region, and we evaluate the conformation entropy explicitly by introducing a transfer matrix.
Using these two models, we evaluate the occurrence probability of critical nucleus as a function of the polymer chain stiffness. We found that the critical nucleus in Model M is easier to occur than in Model S because, for semi-flexible chains, the nucleus in Model M can grow by adding a new polymer chain into the nucleus rather than to diminish the loop and tail parts as in the case of Model S.

\section{Introduction}\label{sec:introduction}
For many years, a lot of researchers were attracted by the isothermal crystallization process of polymers, which is composed of multi-step ordering processes\cite{Strobl_mesomorphic, Chuang, Jheng}.
Especially, early stage of the crystallization has been extensively investigated both experimentally\cite{Chuang, Jheng, imai1, Z-G-Wang, Z-G-Wang_PE, Panine} and theoretically\cite{LH, Waheed, Anwar_2013, Welch}.

In the early stage, some scenarios of ordering mechanisms have been proposed based on experimental results, for example, spinodal-assisted crystallization\cite{imai1, imai2, imai3}, nucleation and growth\cite{Z-G-Wang, Jheng} and appearance of a mesomorphic phase\cite{Strobl_mesomorphic, Konishi}.
The mesomorphic phase, {which} is an intermediate state between the liquid and the solid phases, has the same symmetry as the liquid crystal.
For example, in the case of poly(butylene-2, 6-naphthalate) (PBN), the structure of the mesomorphic phase has a smectic periodicity which is composed of a one dimensional {periodic} order along a {certain direction} and a liquid-{like} structure in the plane perpendicular to the {direction}\cite{Konishi}.
In our previous research\cite{Yokota_induction}, we focused on the spinodal-assisted crystallization, where the spinodal decomposition during induction period of the crystallization is suggested based on X-ray scattering experiments on polyethylene terephthalate\cite{imai1, imai2, imai3}.
This mechanism is supported by a molecular dynamics simulation of united atom model\cite{Gee}.

On the other hand, counter examples against the spinodal-assisted crystallization were reported based on the X-ray scattering experiments on polyethylene and on polypropylene\cite{Z-G-Wang, Z-G-Wang_PE, Panine}.
In our previous research, by constructing a theoretical model, we obtained a criterion of the spinodal decomposition where the stiffness of the polymer chain and the strength of the nematic interaction determine the condition of the spinodal decomposition to occur.

In the present research, we focus on the time regime after the induction period, {\it i.e.}, the time regime where a critical nucleus appears.
Here, the critical nucleus is defined as the minimum-size nucleus which can grow stably.

Recently, the existence of the nuclei in polymer crystallization was directly confirmed by small angle and wide angle X-ray scattering (SAXS and WAXS) experiments, Fourier transform infrared spectroscopy (FTIR) and so on\cite{Chuang, Jheng}.
Moreover, these experimental results were supported by recent particle simulations using molecular dynamics and Monte Carlo techniques\cite{Anwar_2013, Anwar_2015}.

Many researchers have tried to uncover the mechanism of such nucleation process by proposing simple but essential models for the polymer crystallization\cite{Anwar_2015}.
One of the most primitive theoretical models is the classical nucleation theory (CNT){\cite{CNT}}, where a competition between the bulk free energy difference and the surface excess free energy determines whether the nucleus grows or shrinks.
In the CNT, by calculating the free energy difference before and after a nucleus appears, we obtain the size and the occurrence probability of the critical nucleus.
A nucleus {shrinks if it is} smaller than the critical size, while a nucleus larger than the critical size grows.
It is noted that the CNT accounts for both of homogeneous and inhomogeneous nucleation processes where the homogeneous nucleation is driven by thermal fluctuation {in a uniform system} while the inhomogeneous nucleation is initiated by the impurity such as a solid wall.
In 1960, Lauritzen and Hoffman applied the CNT for {both of} homogeneous {and inhomogeneous} nucleations {in polymer crystallization}, where they assumed an anisotropic shape such as a cylindrical shape for the nucleus\cite{LH}.
{In the homogeneous nucleation process,} the free energy difference before and after the nucleation, $\Delta f$, of an isolated nucleus is given by
\begin{align}
\Delta f(r, l)=-\pi r^2 l \Delta \mu
+2\pi r^2 \sigma_{\rm t}+2\pi r l \sigma_{\rm s},\label{eqn:LH_free_energy} 
\end{align}
where $r$ and $l$ are the radius of the top surface {and the height} of the cylindrical shaped nucleus.
Moreover, $\Delta \mu, \sigma_{\rm s}$ and $\sigma_{\rm t}$ are the bulk energy difference per volume between liquid and solid, the surface tension of the side surface and, that of  the top/bottom surfaces. 
We note that $\sigma_{\rm t}$ implicitly includes the effect of the conformation entropy of the chains.

Let us consider a free energy landscape defined by a curved surface $\Delta f(r, l)$ on the $(r, l)$ plane.
We can easily confirm that this free energy landscape has a saddle point at $(r^\ast, l^\ast)$, which specifies the size of the critical nucleus\cite{LH}.
By using the value of $\Delta f(r^\ast, l^\ast)$, we can evaluate the activation energy of the critical nucleus which determines its occurrence probability.
{Here}, it should be noted that, {among the two types of studies performed by Lauritzen and Hoffman, } {only} the CNT for the inhomogeneous nucleation of polymer crystallization is {called} Lauritzen-Hoffman (LH) theory\cite{LH, Huang_Chuang_PTT}.
In LH theory, the nucleation is assumed to start from a solid wall, where the chains in the nucleus are aligned parallel to the solid wall.
In this case, the area of the side surface of the nucleus is less than the one in the case of the homogeneous nucleation, which leads to a change in the contribution from the surface tension of the side surface in eqn.~(\ref{eqn:LH_free_energy}).
LH theory was successfully used in the analysis of the growth rate of the spherulite, where the values of the surface tensions are regarded as fitting parameters.

Despite of the importance of the critical size of the nucleus in the theoretical studies, it is difficult to measure it experimentally due to the limitation of the spatial resolution of experiments.
Then, computer simulation is a powerful tool to investigate the characteristics of the critical nucleus.
In fact, many computer simulations unveiled the dynamics of the nucleation and the {structural} characteristics of the nucleus\cite{Anwar_2013, Welch}.
In these studies, however, the sizes of the critical nuclei have ambiguity because the crystalline region is not well-defined\cite{Anwar_2015, Sommer_Luo_2010}.
The crystalline region is usually defined by introducing threshold values of orientational order parameters which describe the interchain and intrachain bond correlations\cite{Welch}.
Here, a problem comes from the fact that the threshold values in many simulation studies are different from each other.
Although Welch reported a technique to evaluate these threshold values autonomously by using machine learning, there still remains a problem on the precision of the calculation\cite{Welch}.

Some simulation studies report{ed} that chain conformation in the nucleus {affects} the condition of the crystallization\cite{Anwar_2015, Yamamoto_2013}.
The nucleation from a large solid wall of a bulk crystal leads to the appearance of a hairpin-like structure of a chain\cite{Yamamoto_2013}.
On the other hand, such a hairpin-like structure does not appear in {the} simulation study on the homogeneous nucleation\cite{Anwar_2015}.
Although Anwar {\it et al.} concluded that the difference between these behavior is due to the condition of the nucleation\cite{Anwar_2015}, the detail of the physical mechanism has not been clarified.
To explain these results of the simulations, we need to incorporate the effect of the conformation entropy into the analytical model such as CNT. 
Such an attempt has first been done by Muthukumar in 2003, where an extension of CNT was proposed for crystallization of a single polymer chain by adding the effect of conformation entropy\cite{Muthukumar}.
In this model, a nucleus is assumed to be composed not only of straight parts but also of tails and loops (see Fig.~\ref{fig:nucleus_with_conformation} (a)), where the tails and the loops are described by bead spring model. 
On the other hand, the effect of the conformation of single chain system is different from the one of the multi-chain system whose crystallization has been one of the main targets in the experiments and the simulations.
In addition {to} the number of chains participating in the nucleus, the chemical detail of the polymer chains is another important factor for the nucleation.
Although results of the experiments and of the simulations depend on the chemical details\cite{Z-G-Wang_PE, Anwar_2013, Z-G-Wang, Panine}, Muthukumar's model cannot {reflect} such a dependence because Muthukumar assumed that the polymers are simple flexible bead-spring chains.

In this paper, we construct nucleation theory which is applicable to both of single chain and multi-chain systems. 
Our model includes a chain stiffness, {which reflects an important chemcial detail of the polymer chain,} through the calculation of the conformation entropy, where the stiffness is described by the persistence length or by the energy difference between trans and gauche conformations.
Here, the chain stiffness affects the important chemical details of the polymer chains in the samples.

This paper is organized as follows.
In the next section, we discuss our theoretical models of nucleation in single-chain and multi-chain systems.
First, we construct a theoretical model for single-chain systems.
Next, a model for multi-chain systems is constructed based on the single-chain model introduced above.
In Sec.~\ref{sec:result_and_discussion}, we compare {these} two models by {discussing} the effect of the stiffness of the polymer chain on the size of the critical nucleus.
We further discuss the relationship between the induction time and the {stiffness of the polymer chain}. 
Finally, we conclude our results in Sec.~\ref{sec:conclusion} 

\section{Model}\label{sec:model4}
\subsection{Nucleation theory for single polymer chain (Model S)}\label{sec:single_polymer_chain}
\begin{figure}[H]
  \begin{center}
    \includegraphics[width=10.0cm]{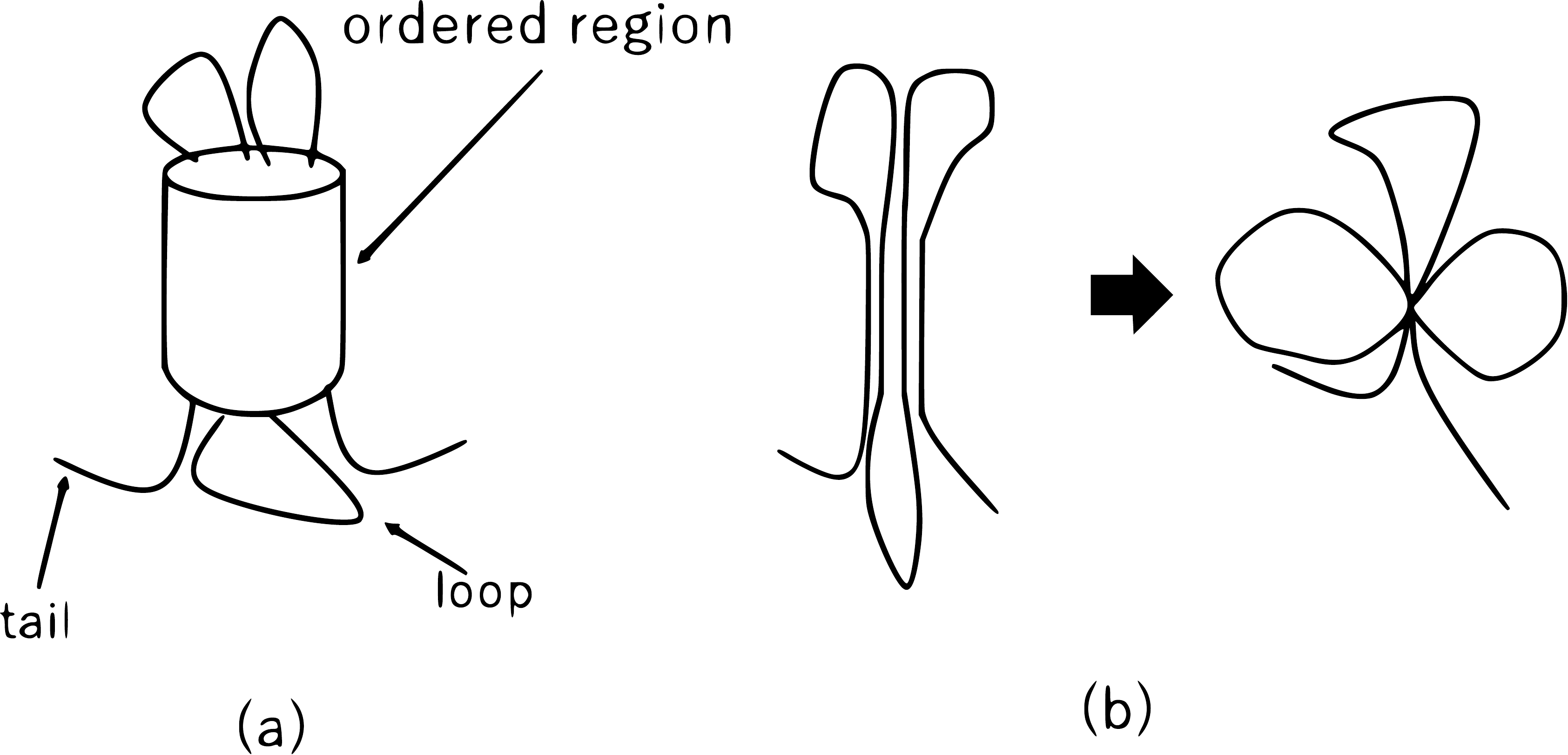}
    \caption{ (a)A schematic picture of Model S of the nucleus.
    The tails and loops are constructed by a semi-flexible chain.
     (b)If the ordered region is negligibly thin, the conformation entropy of the tails and of the loops are approximated as the one of the flower micelle {because} the number of micro states of the ordered region is just unity.
    }
    \label{fig:nucleus_with_conformation}
  \end{center}
\end{figure}

In this subsection, we construct a theoretical model of nucleation of a single polymer chain (Model S) which serves as a basic component of the multi-chain model discussed later.
We calculate the free energy difference before and after the nucleation takes place, where {we assume that} the nucleus is formed by a single semi-flexible polymer chain.
{We also assume that} the shape of the nucleus is as shown in Fig.~\ref{fig:nucleus_with_conformation} (a).
This nucleus is composed of straight parts in the ordered region, {and tails and loops outside of the ordered region.}
We assume that the {two} ends of each loop are on the same side of the ordered region.
The free energy difference $\Delta f$ is evaluated as a function of the number of loops $\alpha$ and the height of the ordered region $m\times b$ where $b$ is the size of a segment of the polymer chain ($l=m\times b$, where $l$ appeared in eqn.~(\ref{eqn:LH_free_energy})).
Then, we obtain
\begin{align}
  \Delta f(\alpha, m)&=- (\alpha+1)m b^3 \Delta \mu
                               +2\sqrt{\pi(\alpha+1)}mb^2 \sigma_{\rm s} + 2(\alpha+1)b^2\sigma_{\rm t}\nonumber \\
  &-k_{\rm B}T
  \ln{\left[Z\left(\alpha, m; \frac{\Delta\varepsilon}{k_{\rm B}T}, N+1; M=1\right)\right]}. \label{eqn:free_energy_single_chain}
\end{align}
{In eqn.~(\ref{eqn:free_energy_single_chain}), $\Delta \mu$ is the bulk free energy difference per unit volume, $\sigma_{\rm s}$ the surface tension of the side surface, $\sigma_{\rm t}$ the surface tension of the top/bottom surfaces, {$\Delta\varepsilon$ the energy difference between trans and gauche conformations (the chain stiffness)}, {$k_{\rm B}$ Boltzman constant}, {$T$ the actual temperature,} $M$ the number of chains participating in the nucleus, $N+1$ the total number of segments in a polymer chain and $Z(\alpha, m; \Delta\varepsilon/(k_{\rm B}T), N+1; M=1)$ the partition function of conformations of the tails and the loops of the single chain.}
{In deriving eqn.~(\ref{eqn:free_energy_single_chain}),} we assumed that the segments are closely packed in the ordered region, which leads to the condition on the total volume of the ordered region {as}
\begin{align}
  {\pi m r^2=(\alpha+1) m b^3}.
\end{align}
The reference value of the free energy is that of a single chain in the free space, which corresponds to the free energy in the liquid state composed of theta solvents and of a single chain with $N+1-m(\alpha+1)$ segments.
The bulk energy difference $\Delta \mu \times b^3$ in eqn.~(\ref{eqn:free_energy_single_chain}) is given by
\begin{align}
  \Delta \mu b^3\simeq \frac{T_{\rm m}^{(0)}-T}{T_{\rm m}^{(0)}}\Delta h, 
\end{align}
where $T_{\rm m}^{(0)}$, $T$ and $\Delta h$ are the equilibrium melting temperature, the actual temperature of the system (thus, $T_{\rm m}^{(0)}-T$ is the degree of supercooling) and the heat of fusion per {segment}, respectively\cite{Strobl}.
We {non-dimensionalize} eqn.~(\ref{eqn:free_energy_single_chain}) by {using} $k_{\rm B}T_{\rm m}^{(0)}$ as
\begin{align}
  \frac{\Delta f(\alpha, m)}{k_{\rm B}T_{\rm m}^{(0)}}
  &=- (\alpha+1)m \frac{T_{\rm m}^{(0)}-T}{T_{\rm m}^{(0)}}
   \frac{\Delta h }{k_{\rm B}T_{\rm m}^{(0)}}\nonumber \\
  & \ \ \ +2\sqrt{\pi(\alpha+1)}m \frac{b^2 \sigma_{\rm s}}{k_{\rm B}T_{\rm m}^{(0)}}
  + 2(\alpha+1)\frac{b^2\sigma_{\rm t}}{k_{\rm B}T_{\rm m}^{(0)}}\nonumber\\
  &-\frac{T}{T_{\rm m}^{(0)}}
  \ln{\left[Z\left(\alpha, m; \frac{\Delta\varepsilon}{k_{\rm B}T_{\rm m}^{(0)}}\frac{T_{\rm m}^{(0)}}{T}, N+1; M=1\right)\right]}\label{eqn:free_energy_single_inter}.
\end{align}
Hereafter, we use the {non-dimensional} parameters where the energy, temperature and length are normalized by $k_{\rm B}T_{\rm m}^{(0)}$, $T_{\rm m}^{(0)}$ and $b$, respectively.
For example, we simply refer to $\Delta h/(k_{\rm B}T_{\rm m}^{(0)})$ {as $\Delta h$}.
Thus, eqn.~(\ref{eqn:free_energy_single_inter}) is rewritten in the following {manner}
\begin{align}
  \Delta f(\alpha, m)
  &=- (\alpha+1)m \left(1-T \right)\Delta h
  +2\sqrt{\pi(\alpha+1)}m \sigma_{\rm s}
  + 2(\alpha+1)\sigma_{\rm t}\nonumber\\
  &-T\ln{\left[Z\left(\alpha, m; \frac{\Delta\varepsilon}{T}, N+1; M=1\right)\right]}.\label{eqn:norm_free_energy_single}
\end{align}
For simplicity, we assume that the ordered region is negligibly thin, therefore the conformations of the tails and the loops of the nucleus are assumed to be the same as those {without the ordered region}.
Under this assumption, our model nucleus is just like a flower micelle as shown in Fig.~\ref{fig:nucleus_with_conformation} (b).
The statistical weight of the conformation is evaluated by using the transfer matrix $\mathcal{T}$ where the polymer chain is regarded as a sequence of rod-like segments\cite{Yokota_induction}.
Hereafter we will refer the rod-like segment as simply `segment' unless otherwise noted.
The orientation vector of each segment is chosen from the 12 {basis vectors of a} face centered cubic lattice $\mbox{\boldmath$e$}^{(\eta)}$ ($\eta=1, 2, \cdots 12$), where the basis vectors are defined as $\mbox{\boldmath$e$}^{(1)}=(b,b,0)/\sqrt{2}, \mbox{\boldmath$e$}^{(2)}=(-b,b,0)/\sqrt{2}, \mbox{\boldmath$e$}^{(3)}=(b,0,b)/\sqrt{2}, \mbox{\boldmath$e$}^{(4)}=(-b,0,b)/\sqrt{2}, \mbox{\boldmath$e$}^{(5)}=(0,b,b)/\sqrt{2}, \mbox{\boldmath$e$}^{(6)}=(0,-b,b)/\sqrt{2}$ and $\mbox{\boldmath$e$}^{(\xi)}=-\mbox{\boldmath$e$}^{(\xi-6)}$ ($\xi=7, 8, \cdots 12$).
The transfer matrix is defined as
\begin{align}
  \mathcal{T}_{\eta\xi}=\begin{cases}
    1 \ \ \
    ({\rm if\ neighboring\ 2\ bonds\ are\ in\ trans\ conformation})\\
    \delta \ \ \
    ({\rm if\ neighboring\ two\ bonds\ are\ in\ gauche\ conformation})\\
    0 \ \ \ ({\rm otherwise})\\
  \end{cases}
  ,\label{eqn:transfermatrix}
\end{align}
where $\eta$ and $\xi$ specify the orientations of 2 consecutive segments, respectively.
Moreover, $1$ and $\delta$ are the statistical weights of the trans and the gauche conformations.
The {value} of $\delta$ is described by using the energy difference between trans and gauche conformations $\Delta \varepsilon$ as
\begin{align}
  \delta=\exp{\left[-\frac{\Delta \varepsilon}{T} \right]}.
\end{align}
We will refer $\Delta \varepsilon$ as simply `gauche energy'.
The persistence length {of the chain} is related to $\Delta \varepsilon$ {through} 
\begin{align}
  l_{\rm p}=\frac{1}{\delta} {\propto \exp \left[\frac{\Delta \varepsilon}{T} \right]}.
\end{align}
By using the transfer matrix, the path integral for {}{a} polymer chain {}{composed of} $N+1$ segments ($N\neq 0$) $Q(0, \eta, \mbox{\boldmath$r$};N, \xi, \mbox{\boldmath$r$}^\prime)$ is defined as follows:
\begin{align}
  Q(0, \eta, \mbox{\boldmath$r$}; N, \xi, \mbox{\boldmath$r$}^\prime)&=
  \int d\mbox{\boldmath$q$} \mathcal{\tilde{T}}_{\eta\xi}^N(\mbox{\boldmath$q$})
  \exp{\left[i \mbox{\boldmath$q$}\cdot 
  \left(\mbox{\boldmath$r$}^\prime-\mbox{\boldmath$r$}\right) \right]}\label{eqn:path_integral},
\end{align}
where
\begin{align}
  \mathcal{\tilde{T}}_{\eta\xi}(\mbox{\boldmath$q$})&=
  \exp{\left[-\frac{i}{2}\mbox{\boldmath$q$}\cdot 
  \mbox{\boldmath$e$}^{(\eta)} \right]}
  \mathcal{T}_{\eta\xi}
  \exp{\left[-\frac{i}{2}\mbox{\boldmath$q$}\cdot 
  \mbox{\boldmath$e$}^{(\xi)} \right]}.
\end{align}
Here $\mbox{\boldmath$r$}$ and $\mbox{\boldmath$q$}$ are the position vector and  the wave number vector.
As shown in \ref{sec:Z_1}, the expression of the statistical weight of conformation of the flower micelle $Z(\alpha, m; \Delta\varepsilon/T, N+1; M=1)$ is written as  
\begin{align}
  &Z\left(\alpha, m; \frac{\Delta\varepsilon}{T}, N+1; M=1\right)\nonumber \\
  &=\frac{1}{12}\frac{1}{\alpha !}
  \sum_{n_0=0}^{N-m(\alpha+1)}\sum_{n_1=0}^{N-m(\alpha+1)}\cdots \sum_{n_{\alpha+1}=0}^{N-m(\alpha+1)}
  \sum_{\eta, \eta^\prime, \eta_0, \eta_1, \cdots , \eta_{\alpha+1}}\nonumber \\
  & \ \ \ \times \mathcal{U}_{\eta\eta_0}(n_0)
  \mathcal{J}_{\eta_0 \eta_1}(n_1)\mathcal{J}_{\eta_1\eta_2}(n_2)
  \cdots\mathcal{J}_{\eta_{\alpha-1}\eta_\alpha}(n_\alpha)
  \mathcal{U}_{\eta_{\alpha}\eta_{\alpha+1}}(n_{\alpha+1})\nonumber\\
  & \ \ \ \times \delta_{0, N-m(\alpha+1)-(n_0+n_1+\cdots n_{\alpha+1})}\label{eqn:conformation_single},
\end{align}
where the prefactor $1/12$ is a normalization factor according to the 12 possible orientations of the initial {segment} and $1/\alpha !$ is the correction of the overcounting of the micro states {for loops}.
$\mathcal{U}_{\eta\xi}(n)$ and $\mathcal{J}_{\eta\xi}(n)$ are the statistical weights of a tail and of a loop, respectively{}{, where both of these are composed of $n$ segments}.
The expressions of these statistical weights are as follows:
\begin{align}
  {\rm tail;}\ \ \ \ \ \mathcal{U}_{\eta\xi}(n)
  &=\frac{1}{(1+4\delta)^{n}}{\left(\mathcal{T}^n\right)_{\eta\xi}} \label{eqn:def_of_tail}, \\
  {\rm loop;} \ \ \ \ \ \mathcal{J}_{\eta\xi}(n)
  &=\frac{1}{(1+4\delta)^{n}}\int d\mbox{\boldmath$q$}
  {\left[\left( \mathcal{\tilde{T}}(\mbox{\boldmath$q$}) \right)^n\right]_{\eta\xi}}\label{eqn:def_of_loop}.
\end{align} 
Using these definitions, eqn.~(\ref{eqn:conformation_single}) is rewritten as
\begin{align}
  &Z\left(\alpha, m; \frac{\Delta\varepsilon}{T}, N; M=1\right)\nonumber \\
  &=\frac{1}{12}\frac{1}{\alpha !}\frac{1}{N-m(\alpha+1)+1}
  \sum_{p=0}^{N-m(\alpha+1)} { \Biggl\{ } 
  \mathcal{\check{U}}_{\eta_0{\eta_1}}(p)
  {\left[ \left( \mathcal{\check{J}}(p) \right)^\alpha \right]_{\eta_1\eta_\alpha} }
  \mathcal{\check{U}}_{\eta_\alpha\eta_{\alpha+1}}(p)  \nonumber \\
  & \times \exp{\left[i\frac{2\pi p(N-m(\alpha+1))}{N-m(\alpha+1)+1} \right]}  { \Biggr\} },
\end{align}
where \ $\check{}$ \ means discrete Fourier transformation with respect to the segment index, for example, 
\begin{align}
  \mathcal{\check{U}}_{\eta\eta_0}(p)=\sum_{n=0}^{N-m(\alpha+1)}\mathcal{U}_{\eta\eta_0}(n)
  \exp{\left[-i\frac{2\pi}{N-m(\alpha+1)+1}pn\right]}.
\end{align}

This Model S gives us the information about the critical nucleus composed of a single polymer chain.
In the next subsection, we construct a theoretical model of the {nucleus} of multi-chain system based on this Model S.
\subsection{Nucleation theory for multi-chain system (Model M)}\label{sec:multi_polymer_chain}
\begin{figure}[H]
  \begin{center}
    \includegraphics[width=6.0cm]{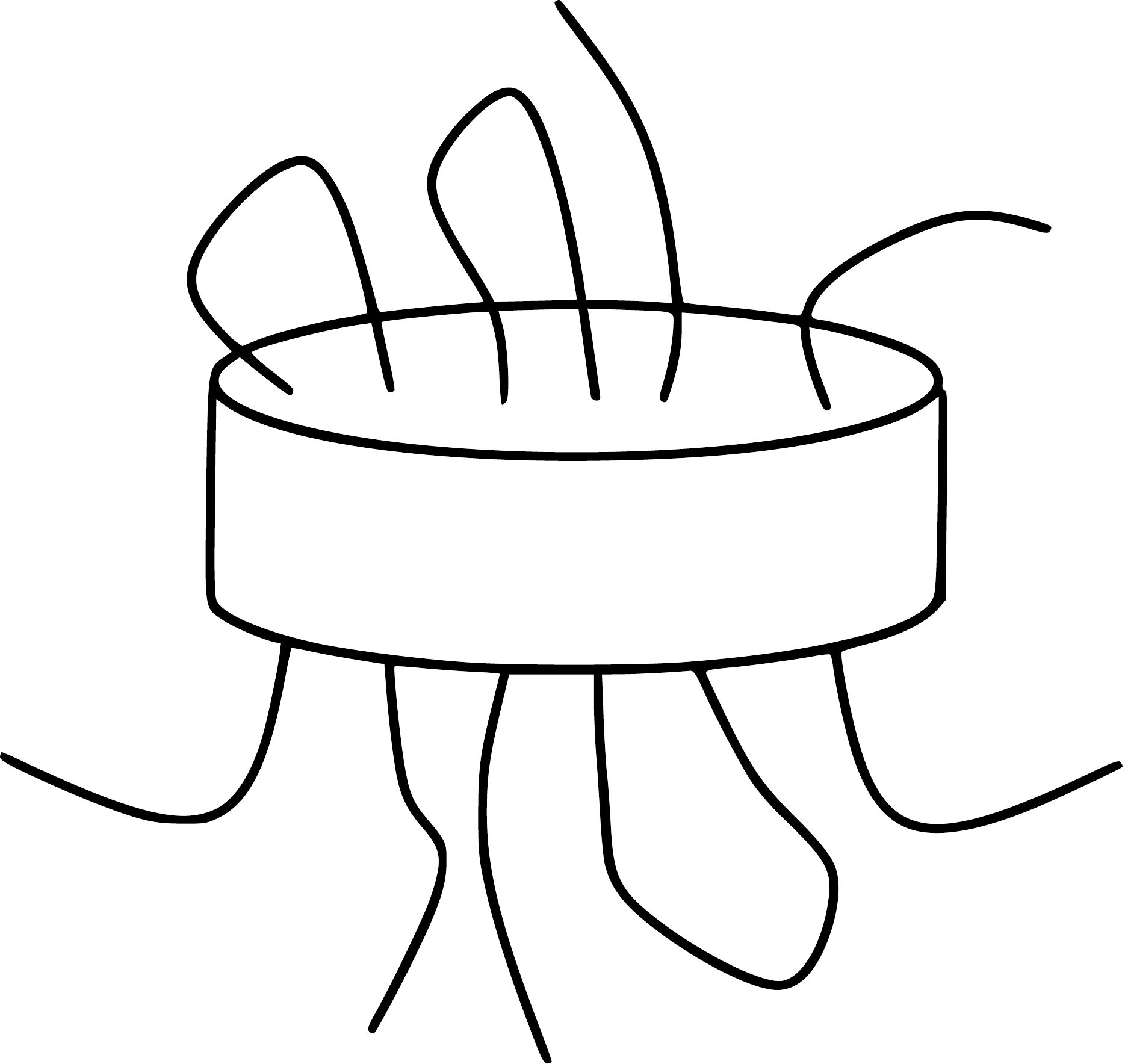}
    \caption{Our model of nucleus (Model M) composed of multi-chains.
    In this example, the number of chains participating in the nucleus $M$ is 3.
    Moreover, the number of loops in the nucleus $\alpha$ is $3$. 
    }
    \label{fig:nucleus_with_conformation_multi}
  \end{center}
\end{figure}

In this subsection, we model the nucleus composed of monodispersed polymer chains which are statistically independent with each other (Model M). 
A schematic picture of the model of nucleus {in} our model M is shown in Fig.~\ref{fig:nucleus_with_conformation_multi}.
In Model M, the number of chains {participating} in the nucleus can fluctuate.
Thus, we should introduce the chemical potential conjugate to the number of chains as an independent variable.
This choice of the variable means that {}{we use} the grand canonical ensemble to evaluate the conformation entropy.
In the case of the nucleus with $\alpha$ loops and with length of ordered region $m\times b$, the total grand potential difference before and after the nucleation is as follows:
\begin{align}
  \Delta \Omega (\alpha, m; \mu_{\rm c}) &=- (\alpha+\langle M\rangle)m b^3 \Delta \mu \nonumber \\
                               &+2\sqrt{\pi (\alpha+\langle M\rangle)}mb^2 \sigma_{\rm s} 
                              + 2(\alpha+\langle M\rangle) b^2\sigma_{\rm t}\nonumber \\
                               &-T\ln{\Xi\left(\alpha, m; \frac{\Delta\varepsilon}{T}, N+1; \mu_{\rm c}\right)},\label{eqn:free_energy_multi_chain}
\end{align} 
where $\mu_{\rm c}$ is {the} chemical potential conjugate to the number of chains and $\Xi(\alpha, m;\Delta\varepsilon/T, N+1;\mu_{\rm c})$ is the grand partition function of {the nucleus including the effect of} the conformation entropy of the tails and loops.
$\langle M \rangle$ is the average number of the chains participating in the nucleus:
\begin{align}
  \langle M \rangle&=-\frac{\partial }{\partial \mu_{\rm c}}
  \left[-\ln{\Xi\left(\alpha, m; \frac{\Delta\varepsilon}{T}, N+1; \mu_{\rm c}\right)} \right] \label{eqn:chemi_pote}.
\end{align} 
{Here we ignore the contribution from CNT terms to the value of $\langle M \rangle$ to simplify the calculation.}

To evaluate $\Xi(\alpha, m; \Delta\varepsilon/T, N+1; \mu_{\rm c})$, {we start from the partition function of the nucleus composed of $M$ chains, where the number of loops and the hight of the nucleus are $\alpha$ and $m$, respectively}.
(For example, in the case of Fig.~\ref{fig:nucleus_with_conformation_multi}, $M=3$ and $\alpha=3$.)
We denote {the} canonical partition function of a nucleus with $M$ chains as $Z (\alpha, m; \Delta\varepsilon/T, N+1; M)$.
As in the case of Model S, the width of the ordered region is also assumed to be negligibly thin {also in the present Model M}.
{Then, we obtain}
\begin{align}
  &Z\left(\alpha, m; \frac{\Delta\varepsilon}{T}, N+1; M\right) \nonumber \\
  &=\frac{\alpha !}{M!}\sum_{\alpha_1, \alpha_2, \cdots, \alpha_{M}=0}^{\alpha}\left\{
   Z\left(\alpha_1, m; \frac{\Delta\varepsilon}{T}, N+1; M=1\right)
   \right. \nonumber \\
  & \ \ \ \times Z\left(\alpha_2, m; \frac{\Delta\varepsilon}{T}, N+1; M=1\right)\times \cdots 
  \times Z\left(\alpha_{M}, m; \frac{\Delta\varepsilon}{T}, N+1; M=1\right)\nonumber\\
  & \ \ \ \times \left. \delta_{0, N\times M-m(\alpha+M)-
  (N-m(\alpha_1+1)+N-m(\alpha_2+1)+\cdots N-m(\alpha_{M}+1))} \right\}\nonumber\\
  &=\frac{\alpha !}{M!}
   \sum_{\alpha_1, \alpha_2, \cdots, \alpha_{M}=0}^{\alpha} \left\{
   Z\left(\alpha_1, m; \frac{\Delta\varepsilon}{T}, N+1; M=1\right)
    \right. \nonumber \\
  & \ \ \ \times Z\left(\alpha_2, m; \frac{\Delta\varepsilon}{T}, N+1; M=1\right) \times \cdots 
  \times  Z\left(\alpha_{M}, m; \frac{\Delta\varepsilon}{T}, N+1; M=1\right)\nonumber\\
  & \ \ \ \times \left.  \delta_{0, \left(\alpha-\sum_{i=1}^{M} \alpha_i\right)} \right\} \\
  &=\frac{\alpha !}{M!}\frac{1}{\alpha+1}\sum_{s=0}^{\alpha}
  \left[ \bar{Z}\left(s, m; \frac{\Delta \varepsilon}{T}, N+1; M=1\right)\right]^{M} \exp{\left[i\frac{2\pi s \alpha}{\alpha+1} \right]},
\end{align}
where \ $\bar{}$ \ means the discrete Fourier transformation with respect to $\alpha$ and, $s$ is the conjugate variable to $\alpha$.
The expression of $\Xi(\alpha, m; \Delta\varepsilon/T, N+1; \mu_{\rm c})$ is as follows:
\begin{align}
  \Xi\left(\alpha, m; \frac{\Delta\varepsilon}{T}, N+1; \mu_{\rm c}\right)
  &=\sum_{M=0}^\infty \exp{\left[\frac{\mu_{\rm c}M}{T}\right]}
  Z\left(\alpha, m ; \frac{\Delta\varepsilon}{T}, N+1; M\right) \label{eqn:Legendre},
\end{align}
{where $\exp{\left[\mu_{\rm c}/T \right]}$ is the so-called `fugacity' }.
By calculating eqn.~(\ref{eqn:free_energy_multi_chain}) for each $\alpha + \langle M \rangle$ and $m$, we can obtain the information about the critical nucleus composed of multiple chains. 
\section{Result and discussion}\label{sec:result_and_discussion}
\subsection{Result of Model S}
In this subsection we discuss the size and the occurrence probability of the critical nucleus in Model S.
In {the} present work, {as} we are interested in the dependence of the effect of the conformation entropy on $\Delta\varepsilon/T$, {we evaluate the free energy difference eqn.~(\ref{eqn:norm_free_energy_single}) when} the parameters $\Delta h$, $\sigma_{\rm s}$ and $\sigma_{\rm t} $ are fixed.

\begin{figure}[H]
  \begin{center}
    \vspace{-20mm}
    \includegraphics[width=13.0cm]{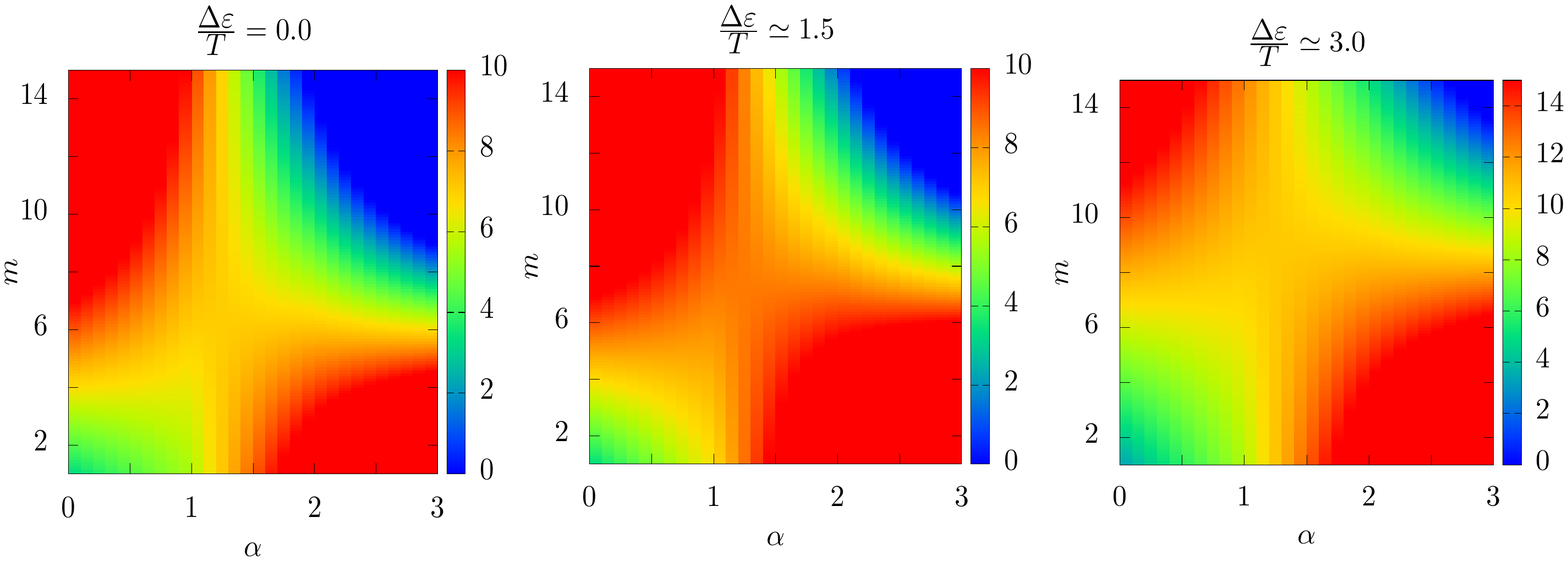}
    \vspace{-20mm}
    \caption{The excess free energy of a nucleus {obtained using Model S} for various values of the gauche energy.
    The vertical axis is the height of the nucleus $m$ 
    and the horizontal axis is the number of loops $\alpha$.
    The parameters are set as $N=127$, {$\Delta h = 23.75$},
   $\sigma_{\rm s}=\sigma_{\rm t}=1.0$ and $T=0.90$.
    }
    \label{fig:single_free_energy_difference_N127}
  \end{center}
\end{figure}
First we show the excess free energy of a nucleus in Fig.~\ref{fig:single_free_energy_difference_N127}.
The size of the critical nucleus is specified by the saddle point {of} the excess free energy $\Delta f(\alpha, m)$ shown in Fig.~\ref{fig:single_free_energy_difference_N127}.
In general, the saddle point {of} a continuous function $g(x, y)$ is obtained by $\partial g/\partial x=\partial g/ \partial y=0$ and by $\lambda_1\times \lambda_2<0$ where $\lambda_1$ and $\lambda_2$ are eigen values of Hessian matrix of $g(x, y)$, {\it i.e.}, 
\begin{align}
    H=\left(
  \begin{array}{cc}
    \frac{\displaystyle \partial^2 g}{\displaystyle \partial x^2} & 
    \frac{\displaystyle \partial^2 g}{\displaystyle \partial x \partial y} \\
    \frac{\displaystyle \partial^2 g}{\displaystyle \partial x \partial y} & 
    \frac{\displaystyle \partial^2 g}{\displaystyle \partial y^2}
  \end{array}
  \right).
\end{align} 
In our model, the variables $\alpha$ and $m$ in $\Delta f$ are discrete values, thus, the saddle point ($\alpha^\ast, m^\ast$) cannot be calculated by using the derivatives of $\Delta f$ with respect to $\alpha$ and $m$.
Instead, the activation energy in our model is defined as the lowest energy barrier among the all paths from $(\alpha, m)=(0, 0)$ to  $(\alpha^\prime, m^\prime)$, where $(\alpha^\prime, m^\prime)$ are the values on the other side of the energy barrier.
The values of {$(\alpha^\ast, m^\ast)$} which specify the activation energy {give} the size {of critical nucleus}.
To obtain the activation energy and the critical size, we {should choose} the lowest energy barrier in the paths from $(\alpha, m)=(0, 0)$ to $(\alpha^\prime, m^\prime)$.
Since $(\alpha^\prime, m^\prime)$ specify the values on the other side of the energy barriers, $(\alpha^\prime, m^\prime)$ satisfy  
\begin{align}
  \begin{cases}
    &\sqrt{(\alpha^\prime)^2+(m^\prime)^2}<L\\
    &\alpha^\prime>0\\
    &m^\prime>0
  \end{cases}.
\end{align}
{Here, $L > 0$ is chosen as an appropriate value that the size of the critical nucleus does not change.}
Here, to search all paths from $(\alpha, m)=(0, 0)$ to $(\alpha^\prime, m^\prime)$, the backtracking algorithm\cite{backtrack} is applied.
{It is noted that, by choosing the small vale of $\epsilon$, we can reduce the computational cost for searching the all paths from $(\alpha, m)$ to $(\alpha^\prime, m^\prime)$.}

\begin{figure}[H]
  \begin{center}
    \vspace{-20mm}
    \includegraphics[width=13.0cm]{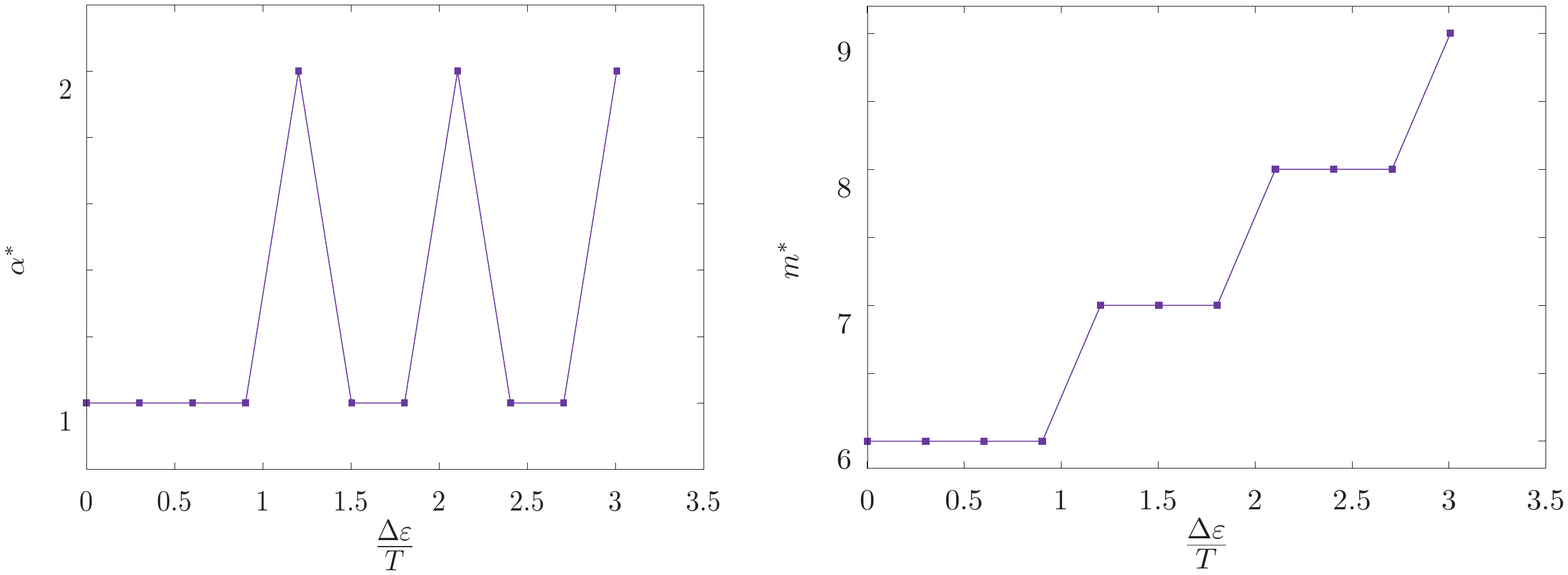}
    \vspace{-20mm}
    \caption{{(left panel)The dependence of $\alpha^\ast$ on $\Delta\varepsilon/T$.}
    {(right panel)}The dependence of $m^\ast$ on $\Delta\varepsilon/T$. 
    {In both panels,} the parameters are the same as the ones in Fig.~\ref{fig:single_free_energy_difference_N127}.
    $m^\ast$ tends to increase with the {gauche energy}, while $\alpha^\ast$ fluctuates.
    }
    \label{fig:single_critical_size}
  \end{center}
\end{figure}

The position of the {critical size} $(\alpha^\ast, m^\ast)$ depends on the gauche energy $\Delta \varepsilon/T$.
In the case of the parameters $\Delta h=23.75$, $T=0.90$ and $\sigma_{\rm s}=\sigma_{\rm t}=1.0$, {the values of} $\alpha^\ast$ and $m^\ast$ depend on the gauche energy as shown in Fig.~\ref{fig:single_critical_size}.
The height of the critical nucleus $m^\ast$ tends to increase with the gauche energy, while $\alpha^\ast$ fluctuates with $\Delta \varepsilon/T$.
The fluctuation of $\alpha^\ast$ is due to the competition between the energy gain of bulk energy difference in the CNT terms and the conformation entropy.
{When $\alpha$ increases, the energy gain from the bulk energy difference increases, while the conformation entropy loss also increases.}
According to the increase in $\Delta \varepsilon/T$, the conformation entropy in eqn.~(\ref{eqn:free_energy_single_chain}) changes.
As a result, $\alpha^\ast$ and $m^\ast$ change with the gauche energy $\Delta \varepsilon/T$.
It should be noted that, in the case of the CNT parameters {that are the} same as those in Fig~\ref{fig:single_critical_size}, the contributions from the CNT terms and the conformation entropy term are {almost the} same when $\alpha$ changes.
These contributions lead to the fluctuation of $\alpha^\ast$.

\begin{figure}[H]
  \begin{center}
    \includegraphics[width=10.0cm]{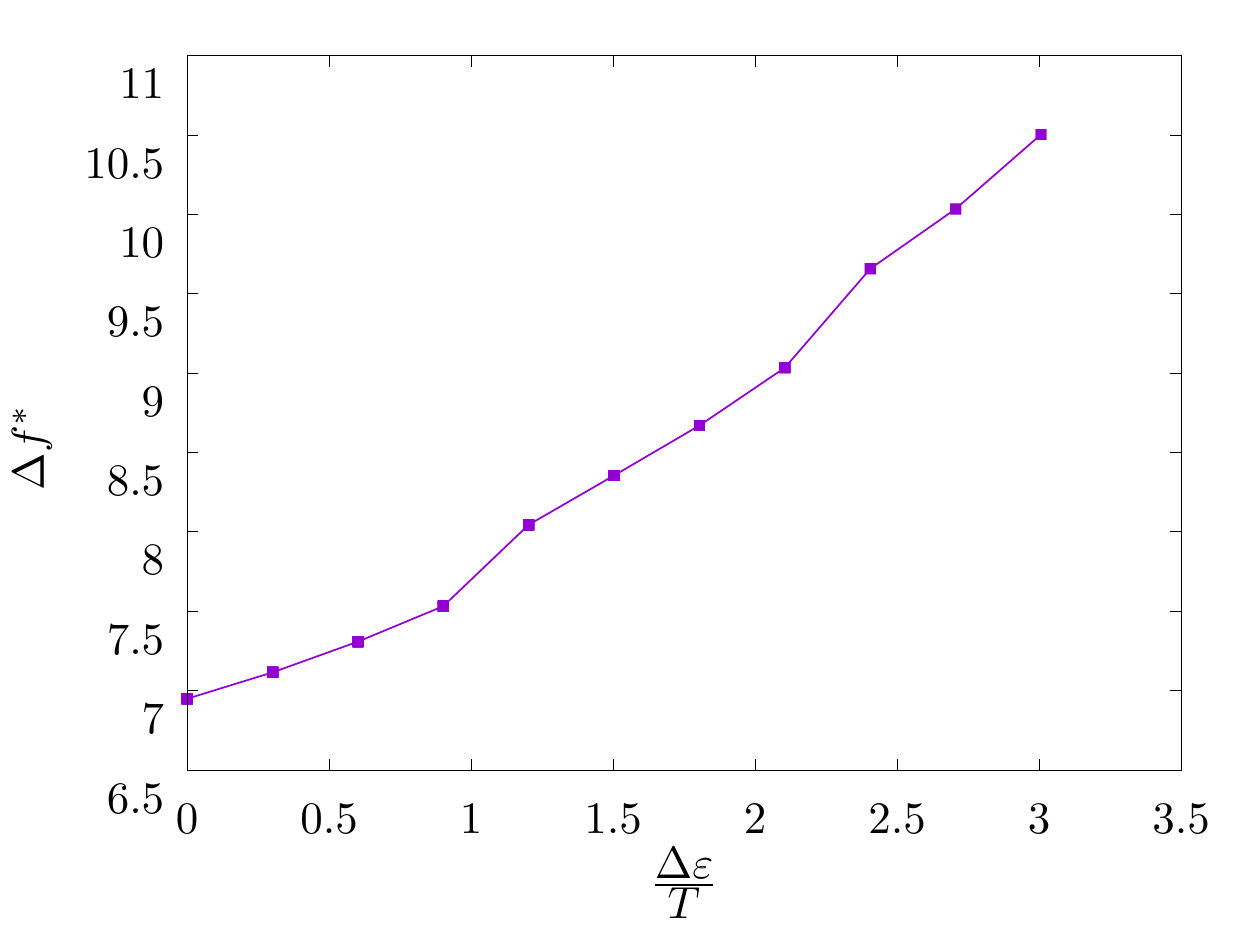}
    \caption{The dependence of $\Delta f^\ast$ on $\Delta\varepsilon/T$.
    The parameters of the CNT are the same as those in Fig.~\ref{fig:single_critical_size}.
    }
    \label{fig:single_activation_energy}
  \end{center}
\end{figure}

We show the behavior of the activation energy $\Delta f^\ast$ as a function of the gauche energy $\Delta\varepsilon/T$ in Fig.~\ref{fig:single_activation_energy}.
%
The behavior of $\Delta f^\ast$ shown in Fig.~\ref{fig:single_activation_energy} is explained {as follows}.
Since the parameters of the CNT are fixed, the activation energy is mainly affected by the conformation {entropy} term in eqn.~(\ref{eqn:free_energy_single_chain}).
When the value of $\Delta \varepsilon/T$ is large, the loss of the conformation {entropy} is relatively large compared with {that} in the case of small $\Delta \varepsilon/T$.
Thus, the activation energy is a monotonically increasing function of $\Delta \varepsilon/T$ {as is shown in Fig.~\ref{fig:single_activation_energy}}.

\begin{figure}[H]
  \begin{center}
    \includegraphics[width=10.0cm]{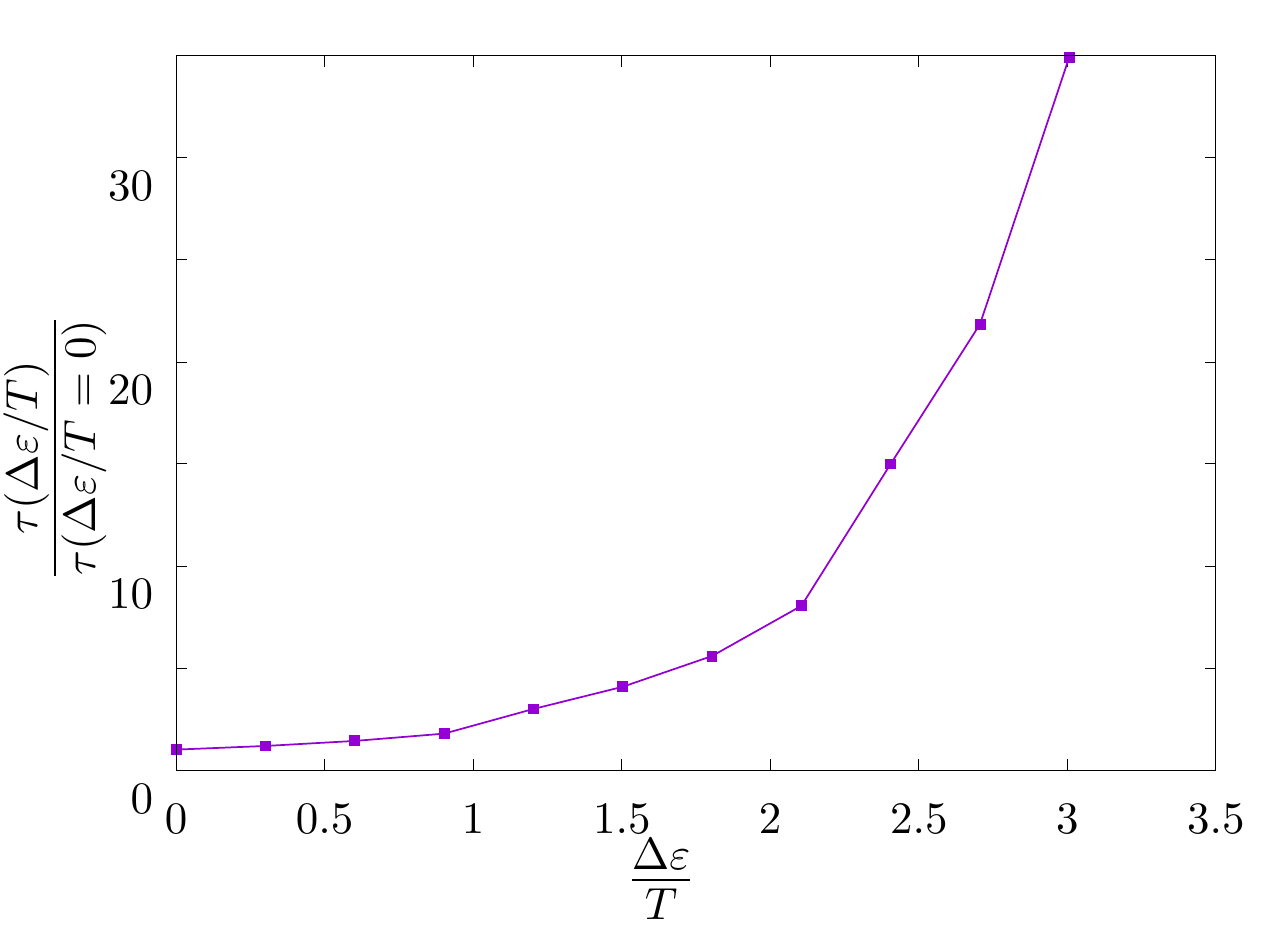}
    \caption{The behavior of {$\frac{\displaystyle \tau(\Delta \varepsilon/T)}{\displaystyle \tau(\Delta\varepsilon/T=0)}$}.
    The parameters of the CNT are the same as the ones in Fig.~\ref{fig:single_critical_size}.
    }
    \label{fig:single_induction}
  \end{center}
\end{figure}

$\Delta f^\ast$ is related to the induction time $\tau$ of the nucleation, which is defined as the period before the critical nucleus is generated, as
\begin{align}
  \tau&=\tau_0 \exp{\left[ \Delta f^\ast \right]},
\end{align}
where $\tau_0$ is the atomistic time scale.
Figure \ref{fig:single_induction} shows the induction time as a function of the gauche energy {for the case} with the same parameters in Fig.~\ref{fig:single_critical_size}.
The dependence of the induction time on the gauche energy is qualitatively the same as the one of the activation energy. 
In experimental studies, the induction time is usually obtained by using DSC {(Differential Scanning Calorimetry)} technique by observation of the latent heat, where the {observed} induction time {corresponds to the generation of several nuclei}.
On the other hand, the induction time in our model is the time by a single critical nucleus appears.
{Although this} induction time is not perfectly {the} same as the one {obtained} with DSC technique, {it qualitatively} corresponds to the experimental {values}.
Moreover, the gauche energy depends on the chemical species of polymers.
The result shown in Fig.~\ref{fig:single_induction} implies the relationship between the observable ``induction time'' and {the} microscopic parameter $\Delta \varepsilon/T$.
\subsection{Result of Model M}\label{sec:result_of_Model_M}
\begin{figure}[H]
  \begin{center}
    \vspace{-20mm}
    \includegraphics[width=12.0cm]{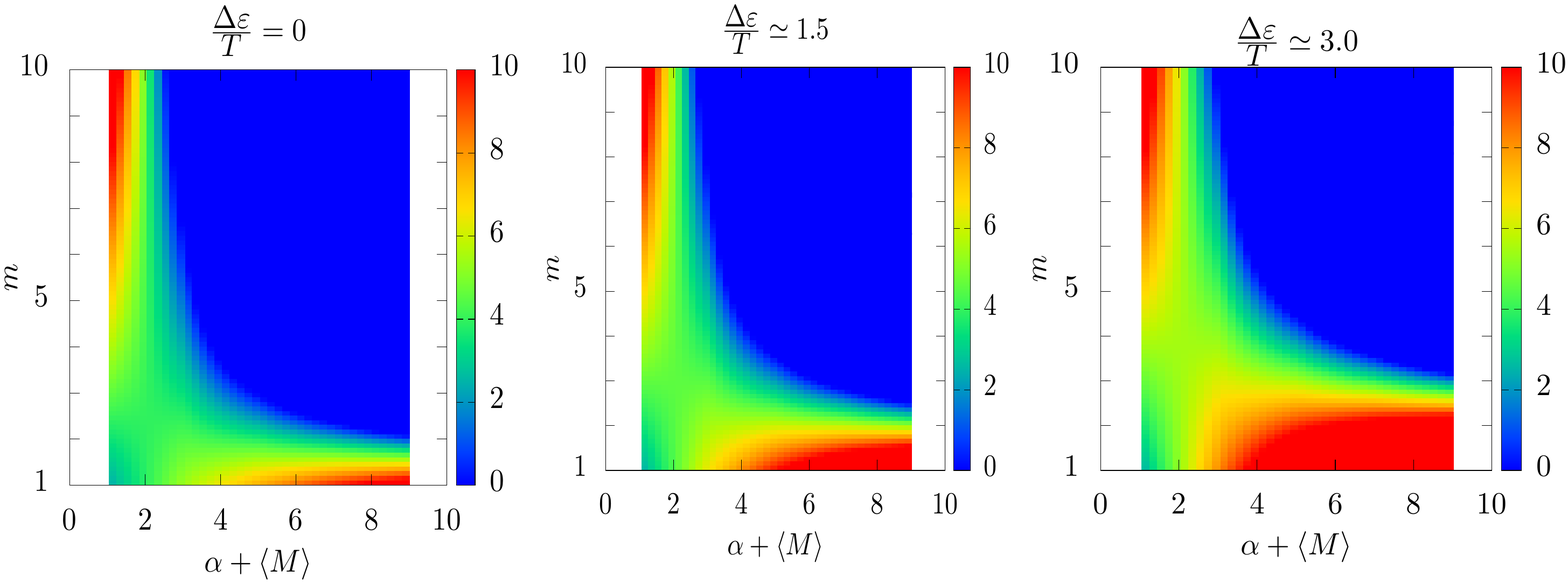}
    \vspace{-19mm}
    \caption{The grand potential difference 
    $\Delta \Omega$ for the various values of
    the gauche energy $\Delta \varepsilon/T$ in Model M.
    The vertical axis is the height of the nucleus $m$ 
    and the horizontal axis is the number of {straight parts in the cylindrical ordered region} $\alpha+\langle M \rangle$.
    The {other} parameters are set as $N=127$, {$\Delta h = 23.75$},
   $\sigma_{\rm s}=\sigma_{\rm t}=1.0$, $T=0.90$ and $\mu_{\rm c}/T=0.0$.
    }
    \label{fig:multi_free_energy_difference_N127}
  \end{center}
\end{figure}

\begin{figure}[H]
  \begin{center}
    \vspace{-20mm}
    \includegraphics[width=13.0cm]{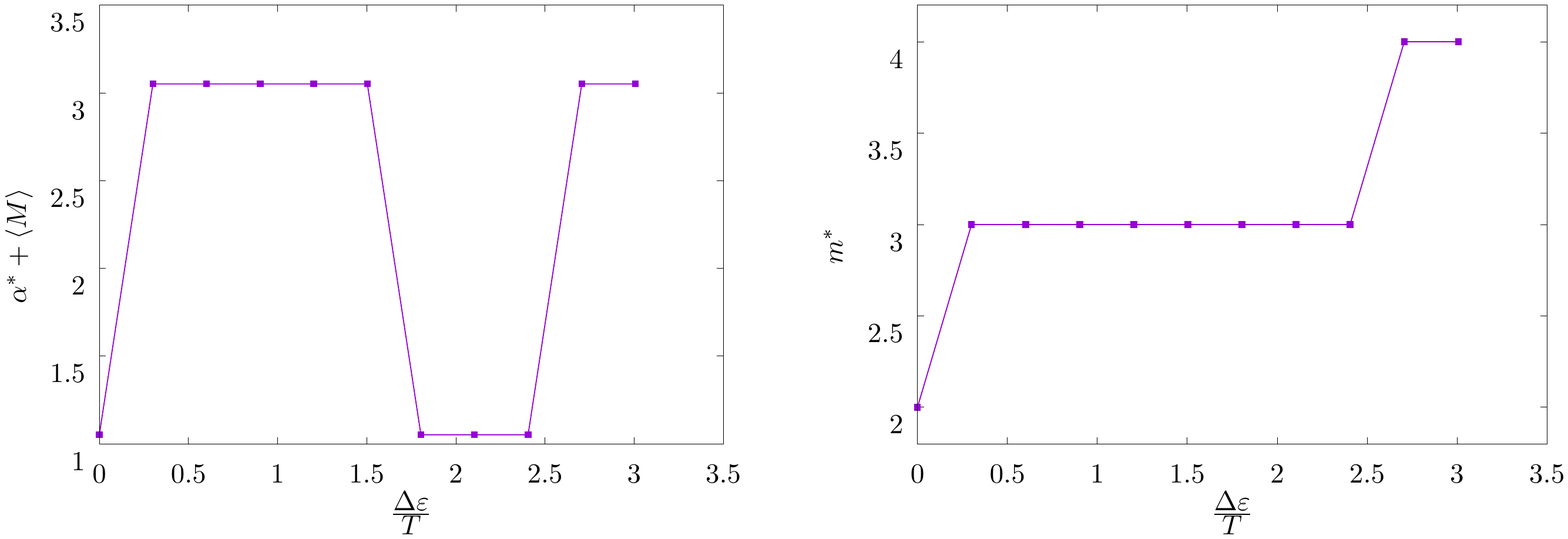}
    \vspace{-30mm}
    \caption{(left panel)The dependence of the number of the straight parts in the critical nucleus $\alpha^\ast+\langle M \rangle$ on the dimensionless gauche energy $\Delta\varepsilon/T$ in Model M.
     (right panel) The dependence of the height of the critical nucleus {}{$m^\ast$} on 
    the dimensionless gauche energy in Model M.
    The parameters are the same as the ones in Fig.~\ref{fig:multi_free_energy_difference_N127}.
    }
    \label{fig:multi_critical_size}
  \end{center}
\end{figure}

In this subsection, the results obtained from Model M are discussed.
First, we discuss the grand potential difference before and after the nucleation, $\Delta \Omega$, for the case $\Delta h=23.75$, $T=0.90$, $\sigma_{\rm s}=1.0$, $\sigma_{\rm t}=1.0$ and $\mu_{\rm c}/T=0.0$ in Fig.~\ref{fig:multi_free_energy_difference_N127}.
The {size of the critical nucleus} $(\alpha^\ast+\langle M \rangle, m^\ast)$ is expected to depend on the gauche energy $\Delta \varepsilon/T$ and the chemical potential $\mu_{\rm c}/T$.
We show the relationship between the critical sizes and the gauche energy in Fig.~\ref{fig:multi_critical_size} in the case of $\mu_{\rm c}/T=0$.
Figure~\ref{fig:multi_critical_size} implies that $\alpha^\ast+\langle M \rangle$ {varies} with gauche energy due to the {change in} the number of loops.
Moreover, the height of the critical nucleus {$m^\ast$} tends to increase with gauche energy.
These dependences of $\alpha^\ast + \langle M \rangle$ and $m^\ast$ on $\Delta \varepsilon/T$ are due to the same mechanism as in Model S.

\begin{figure}[H]
  \begin{center}
    \includegraphics[width=10.0cm]{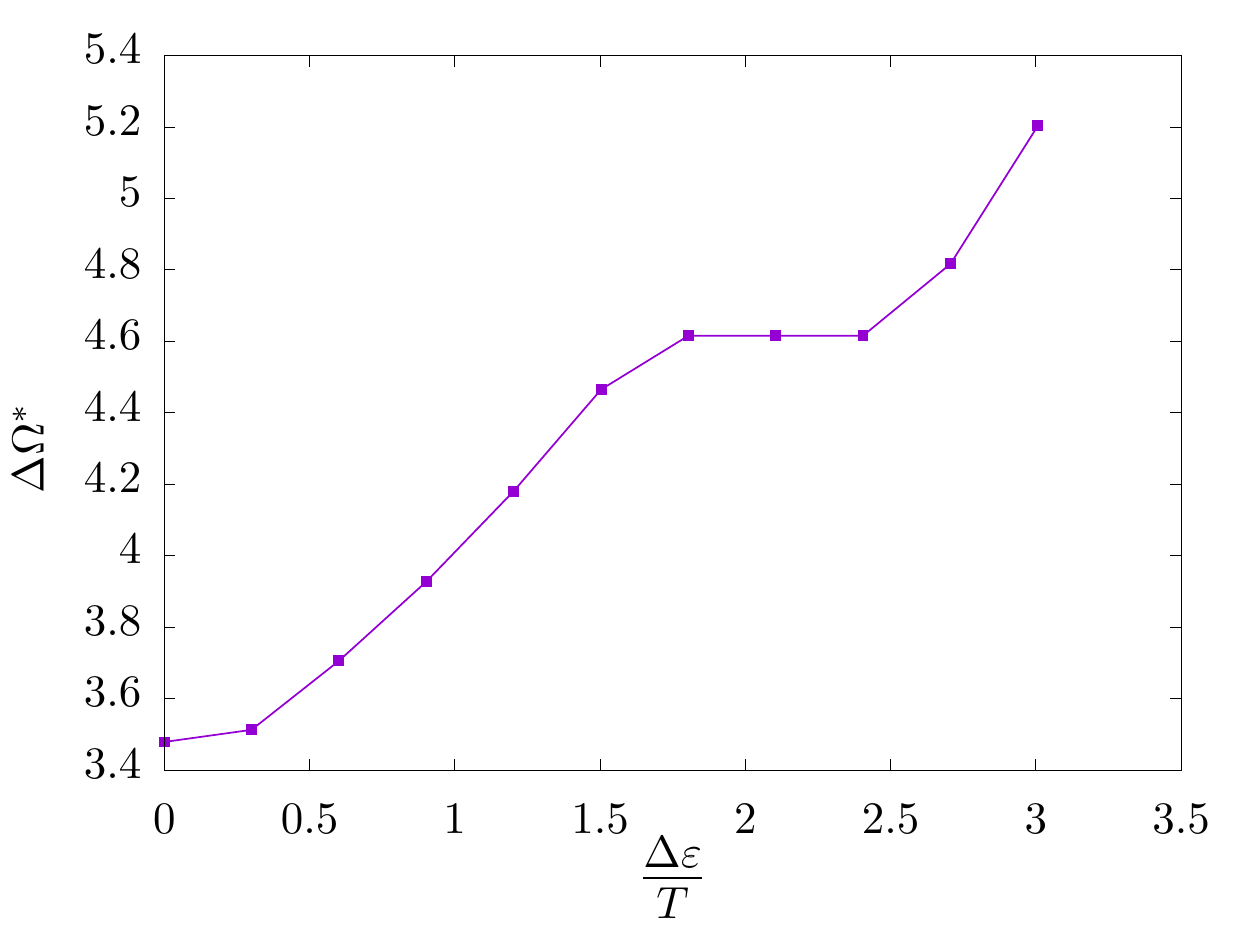}
    \caption{The dependence of the activation energy
    $\Delta \Omega^\ast$ on the 
    gauche energy $\Delta \varepsilon/T$ in Model M.
    The vertical axis is $\Delta \Omega^\ast$ 
    and the horizontal axis is the $\Delta\varepsilon/T$.
    The parameters are the same as the ones in Fig.~\ref{fig:multi_free_energy_difference_N127}.
    }
    \label{fig:multi_activation_energy}
  \end{center}
\end{figure}

We show the dependence of the activation energy {of} Model M on the gauche energy in Fig.~\ref{fig:multi_activation_energy}.
The plateau of $\Delta \Omega^\ast$ corresponds to $\alpha^\ast=0$ $(\alpha^\ast+\langle M \rangle \simeq 1)$. 
It is noted that, as shown in Figs.~\ref{fig:single_activation_energy} and \ref{fig:multi_activation_energy}, $\Delta \Omega^\ast$ in Model M is smaller than $\Delta f^\ast$ in Model S, which is due to the competition between the CNT terms and the conformation entropy.
For example, the conformation entropy in Model M is larger than the one in Model S because the chain does not necessarily generate a {high-energy hairpin} loop in Model M.

\begin{figure}[t]
  \begin{center}
    \vspace{-10mm}
    \includegraphics[width=12.0cm]{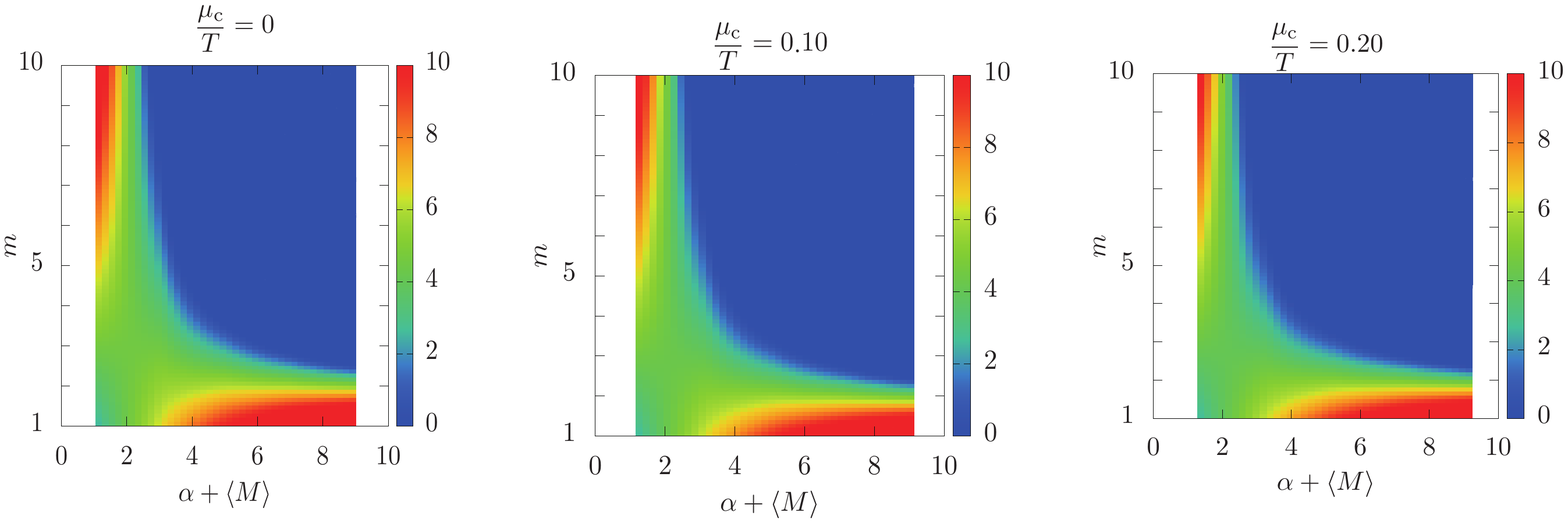}
     \vspace{-25mm}
    \caption{The grand potential difference {for various values of $\mu_{\rm c}/T$}.
    The parameters of classical nucleation terms are
    the same as the ones in Fig.~\ref{fig:multi_free_energy_difference_N127} and $\Delta\varepsilon/T\simeq 1.5$.
    }
    \label{fig:multi_free_energy_difference_mu}
  \end{center}
\end{figure}

\begin{figure}[H]
  \begin{center}
    \vspace{-20mm}
    \includegraphics[width=12.0cm]{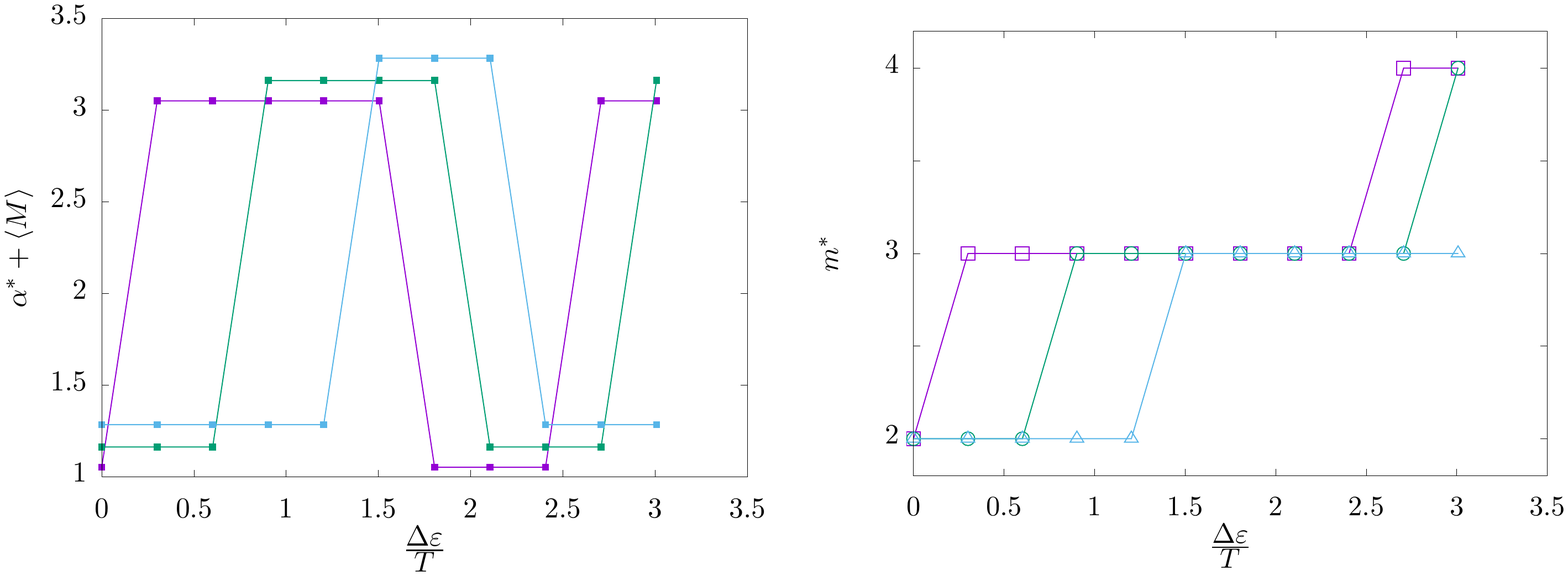}
    \vspace{-20mm}
    \caption{{(left panel)}The dependence of the number of straight parts of the critical nucleus $\alpha^\ast+\langle M \rangle$ on $\Delta\varepsilon/T$ for 
   $\mu_{\rm c}/T=0.00$ (purple), 
    $0.10$ (green) and
    $0.20$ (blue) in Model M {and} (right panel)the dependence of the critical nucleus $m^\ast$.
    The parameters of CNT are the same as the ones in Fig.~\ref{fig:multi_free_energy_difference_N127}.
    }
    \label{fig:multi_critical_size_mu}
  \end{center}
\end{figure}

\begin{figure}[t]
  \begin{center}
    \includegraphics[width=10.0cm]{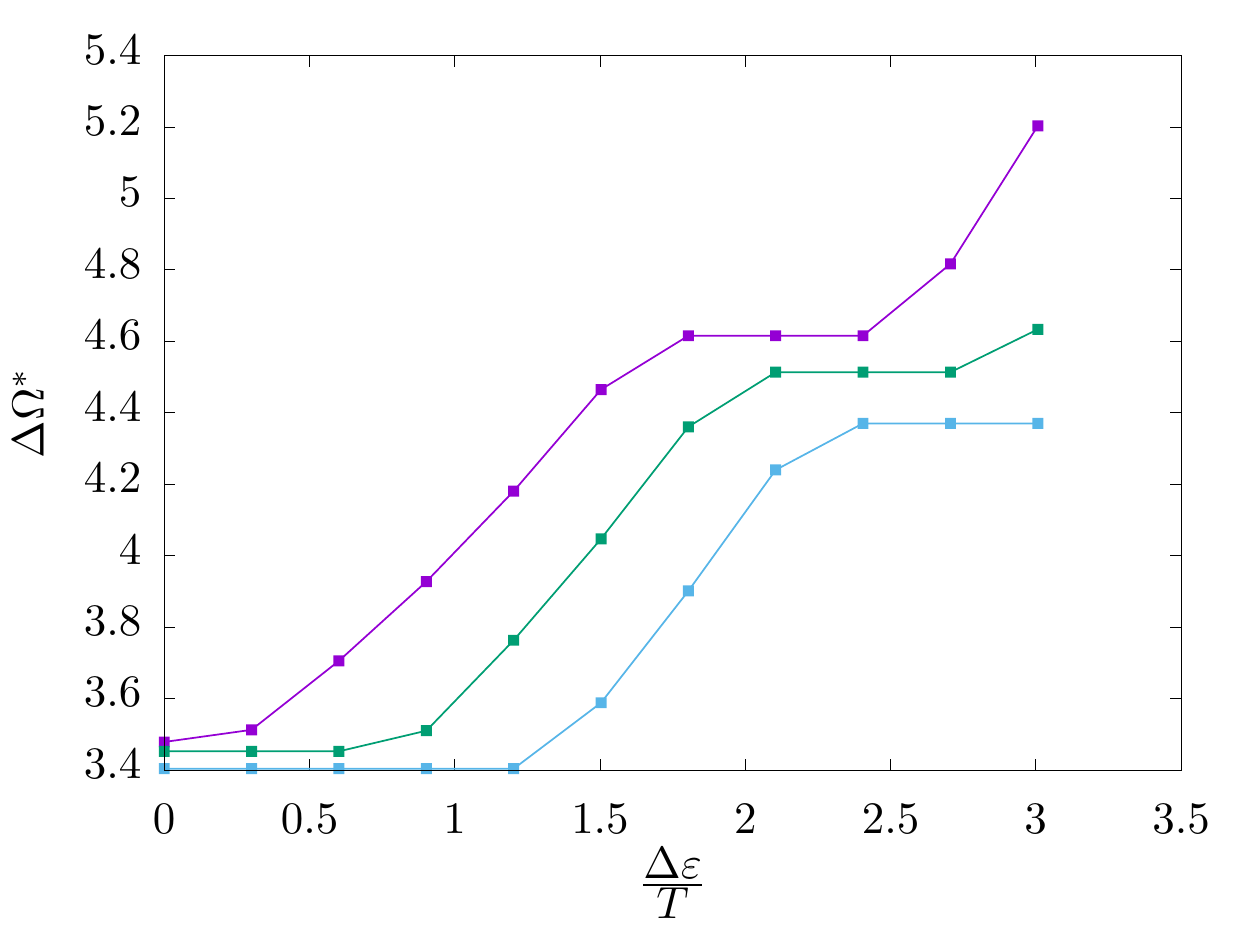}
    \caption{The dependence of the activation energy $\Delta\Omega^\ast$ on
    the dimensionless gauche energy $\Delta\varepsilon/T$ for 
    $\mu_{\rm c}/T=0.00$ ({purple}), 
    $0.10$ ({green}) and
    $0.20$ ({blue}) in Model M.
    The parameters of CNT are the same as the ones in Fig.~\ref{fig:multi_free_energy_difference_N127}.
    }
    \label{fig:multi_activation_energy_mu}
  \end{center}
\end{figure}

Next, we discuss the effect of the chemical potential $\mu_{\rm c}/T$.
We show {in Fig.~\ref{fig:multi_free_energy_difference_mu}} the grand potential difference with the chemical potential where $\Delta\varepsilon/T\simeq 1.5$.
The activation energy in Model M is expected to depend on $\mu_{\rm c}/T$ since $\langle M \rangle$ increases with $\mu_{\rm c}/T$ as is shown in \ref{sec:M_and_mu}.
The increase in $\langle M \rangle$ with $\mu_{\rm c}/T$ is explained by the contribution from the fugacity in eqn.~(\ref{eqn:Legendre}).  
As a large value of $\mu_{\rm c}/T$ leads to a large value of $\alpha+\langle M \rangle$ (see Fig.~\ref{fig:multi_critical_size_mu}), the decrease in bulk free energy term leads to a decrease in the activation energy (see Fig.~\ref{fig:multi_activation_energy_mu}).

The behavior of the size of the critical nucleus for each $\mu_{\rm c}/T$ is shown in Fig.~\ref{fig:multi_critical_size_mu}.
In the {left panel} of Fig.~\ref{fig:multi_critical_size_mu}, $\alpha^\ast + \langle M \rangle$ {varies} with $\Delta \varepsilon/T$ due to the fluctuation of the number of loops.
On the other hand, $\alpha^\ast+\langle M\rangle$ increases with $\mu_{\rm c}/T$ {as long as $\alpha^\ast$ is constant (for example, see the value of $\alpha^\ast+\langle M \rangle$ at $\Delta \varepsilon/T\simeq 1.5$).}
The contribution to the increase in $\alpha^\ast+\langle M \rangle$ from $\langle M \rangle$ is dominant, which means that $\mu_{\rm c}/T$ can be treated as the driving force for a polymer chain to participate in the nucleus.
The gauche energy {which changes $m^\ast$} increases with the chemical potential $\mu_{\rm c}/T$ {as shown in the right panel of Fig.~\ref{fig:multi_critical_size_mu}}.
The behaviors of the activation energy for each $\mu_{\rm c}/T$ are shown in Fig.~\ref{fig:multi_activation_energy_mu}.
{According to Fig.~\ref{fig:multi_activation_energy_mu},} we recognize that $\Delta \Omega^\ast$ monotonically decreases with $\mu_{\rm c}/T$.

In Model M, the size {of the critical nucleus} is determined by $\Delta \varepsilon/T$ and by $\mu_{\rm c}/T$, while in Model S, the critical size is determined only by $\Delta \varepsilon/T$.
\section{Conclusion}\label{sec:conclusion}
We extended the classical nucleation theory (CNT) which does not contain the conformation entropy and the information of the number of chains participating in the nucleus.
We proposed 2 theoretical models of nucleation behavior.
One is the model for a nucleus composed of a single chain (Model S), and the other is the model for a nucleus composed of multiple chains (Model M).
In these models, the nucleus is composed of tails, loops and an ordered region whose shape is assumed to be a cylinder, where the width of cylindrical ordered region is assumed to be negligibly thin. 
By using these models, we discuss the behavior of the critical nucleus in terms of the chain stiffness and of the number of chains participating in the nucleus which are not {taken into account} in Muthukumar's model\cite{Muthukumar}.

In Model S, we obtain the information about the critical nucleus by calculating the free energy difference before and after the nucleation.
The size of the critical nucleus is determined by the competition between the CNT terms and the conformation entropy term.
In our model, the number of loops $\alpha$ and the height of the ordered region $m$ are discrete values. Thus, the critical size and the activation energy cannot be determined by using derivative of the free energy difference with respect to $\alpha$ and $m$.
Then, the activation energy is defined by the lowest free energy among all paths from $(\alpha, m)=(0, 0)$ to $(\alpha^\prime, m^\prime)$ where the latter is the value on the other side of the energy barrier.
Moreover, the size of the critical nucleus is defined as the values of the number of loops and the height of the nucleus which correspond to the saddle point.
The activation energy is a monotonically increasing function of the chain stiffness.
This result is interpreted by using the fact that the conformation entropy loss increases with the chain stiffness.

In Model M, we introduced the chemical potential $\mu_{\rm c}$ which is conjugate to the number of chains participating in the nucleus.
The grand potential difference discussed in Model M for $\mu_{\rm c}/T=0.0$ is smaller than the free energy difference evaluated in Model S.
This means that the critical nucleus can be formed {more} {easily} in the multi-chain system than in the single chain system due to the {difference in the} conformation entropy loss associated with the {creation of} loops.

In Model M, the size of the ordered region is proportional to the average of the number of chains participating in the nucleus $\langle M \rangle$.
As the chemical potential plays a role of a driving force for a polymer chain to participate in the nucleus, $\langle M \rangle$ is an increasing function of the chemical potential.
Thus, the high $\mu_{\rm c}$ leads to a large radius of the cylindrical ordered region despite of {our} assumption {that} the size of the ordered region is negligibly small.
Therefore, {when} we study the crystallization process for high $\mu_{\rm c}$, we cannot approximate the conformation entropy of the nucleus as the one in free space due to the effect {of the excluded volume} of the ordered region.

The number of chains participating in the nucleus is determined by {the following} 2 contributions;
The first one is the contribution from the competition between surface tensions and bulk energy difference which is controlled by the temperature.
The other contribution is the chemical potential conjugate to the number of chains.
If the state of the nucleus is specified, e.g. the size of the ordered region, $\langle M \rangle$ is obtained from the derivative of the grand potential difference with respect to $\mu_{\rm c}$.
\section*{Acknowledgement}
This work is partially supported by the Grant-in-Aid for Scientific Research (Grant Number 26287096 {and} {16K13844}) from The Ministry of Education, Culture, Sports, Science and Technology(MEXT), Japan.

\newpage
\appendix
\def\thesection{Appendix \Alph{section}}

\section{Derivation of {\large{$Z(\alpha, m; \Delta\varepsilon/T, N+1; M=1)$}}}\label{sec:Z_1}
The conformation of a flower micelle is calculated by using the path integral $Q$ defined in eqn.~(\ref{eqn:path_integral}) {as}
\begin{align}
  &Z_1(N+1-m(\alpha+1)) \nonumber \\
  &=\int d{\mbox{\boldmath$r$}}d{\mbox{\boldmath$x$}}
  d{\mbox{\boldmath$r$}^\prime}
  \sum_{n_0=0}^{N-m(\alpha+1)}\sum_{n_1=0}^{N-m(\alpha+1)}
  \cdots \sum_{n_{\alpha+1}=0}^{N-m(\alpha+1)}
  \sum_{\eta, \eta^\prime, \eta_0, \eta_1, \cdots \eta_{\alpha+1}}\nonumber \\
  &\ \ \ \times Q(0, \eta, {\mbox{\boldmath$r$}};n_0, \eta_0, {\mbox{\boldmath$x$}})\nonumber\\
  &\ \ \ \times Q(n_0, \eta_0, {\mbox{\boldmath$x$}} ; n_0+n_1, \eta_1, {\mbox{\boldmath$x$}})
  \cdots
  Q(\sum_{i=0}^{\alpha-1}n_i, \eta_{\alpha-1}, {\mbox{\boldmath$x$}} ; \sum_{i=0}^{\alpha}n_i, \eta_{\alpha},   {\mbox{\boldmath$x$}})\nonumber\\
  &\ \ \ \times Q(\sum_{i=0}^{\alpha}n_i, \eta_{\alpha}, {\mbox{\boldmath$x$}} ; \sum_{i=0}^{\alpha+1}n_i, \eta_{\alpha+1}, {\mbox{\boldmath$r$}^\prime})\nonumber \\
  &\ \ \ \times \delta_{0, N-m(\alpha+1)-(n_0+n_1+\cdots n_{\alpha+1})}\label{eqn:starting_point_of_Z1},
\end{align}
where $\eta, \eta^\prime$ and $\eta_i$ are indices of the segment orientation.
{Variables} $\mbox{\boldmath$r$}$ and $\mbox{\boldmath$r$}^\prime$ are the positions of {two} ends of {the} tails, and $\mbox{\boldmath$x$}$ is the position of the {branching point} of the flower micelle.
By using Fourier transformation with respect to the positions of segments, we derive
\footnotesize
\begin{align}
&Z_1(N+1-m(\alpha+1)) \nonumber \\
&=\frac{1}{N-m(\alpha+1)+1}\frac{1}{12\left(t+4g \right)^{N-m(\alpha+1)}}\nonumber\\
&\ \ \ \times \int d\mbox{\boldmath$r$}d\mbox{\boldmath$x$}
d\mbox{\boldmath$r$}^\prime \nonumber \\
&\ \ \ \times \sum_{n_0=0}^{N-m(\alpha+1)}\sum_{n_1=0}^{N-m(\alpha+1)}
\cdots \sum_{n_{\alpha+1}=0}^{N-m(\alpha+1)}
\sum_{\eta, \eta_0, \eta_1, \cdots \eta_{\alpha+1}}
\int d\mbox{\boldmath$q$}_0\int d\mbox{\boldmath$q$}_1\cdots 
\int d\mbox{\boldmath$q$}_{\alpha+1}\nonumber \\
&\ \ \  \times \left[ \left( \mathcal{\tilde{T}}({\mbox{\boldmath$q$}_0}) \right)^{n_0}\right]_{\eta\eta_0}
\left[ \left( \mathcal{\tilde{T}}({\mbox{\boldmath$q$}_1})\right)^{n_1}\right]_{\eta_0\eta_1}
\left[ \left( \mathcal{\tilde{T}}({\mbox{\boldmath$q$}_2})\right)^{n_2}\right]_{\eta_1\eta_2} 
\cdots
\left[ \left( \mathcal{\tilde{T}}({\mbox{\boldmath$q$}_\alpha})\right)^{n_{\alpha-1}}\right]_{\eta_{\alpha-1}\eta_\alpha}
\left[ \left( \mathcal{\tilde{T}}({\mbox{\boldmath$q$}_{\alpha+1}})\right)^{n_{\alpha+1}}\right]_{\eta_\alpha\eta_{\alpha+1}}
\nonumber \\
&\ \ \  \times \exp{\left[i{\mbox{\boldmath$q$}_0}\cdot 
({\mbox{\boldmath$x$}}-{\mbox{\boldmath$r$}}) \right]}
\exp{\left[i{\mbox{\boldmath$q$}_{\alpha+1}}\cdot 
({\mbox{\boldmath$r$}^\prime}
-{\mbox{\boldmath$x$}} )\right]}\nonumber\\
&\ \ \ \times \delta_{0, N-m(\alpha+1)-(n_0+n_1+\cdots n_{\alpha+1})}\\
&=\frac{1}{N-m(\alpha+1)+1}\frac{1}{12\left(t+4g \right)^{N-m(\alpha+1)}}\nonumber \\
&\ \ \ \times \sum_{n_0=0}^{N-m(\alpha+1)}\sum_{n_1=0}^{N-m(\alpha+1)}
\cdots \sum_{n_{\alpha+1}=0}^{N-m(\alpha+1)}
\sum_{\eta, \eta_0, \eta_1, \cdots \eta_{\alpha+1}}
\int d\mbox{\boldmath$q$}_1\int d\mbox{\boldmath$q$}_2\cdots 
\int d\mbox{\boldmath$q$}_{\alpha}\nonumber \\
&\ \ \ \times \left(\mathcal{T}^{n_0}\right)_{\eta\eta_0}
  \left[ \left( \mathcal{\tilde{T}}({\mbox{\boldmath$q$}_1})\right)^{n_1}\right]_{\eta_0\eta_1}
  \left[ \left( \mathcal{\tilde{T}}({\mbox{\boldmath$q$}_2})\right)^{n_2}\right]_{\eta_1\eta_2} 
  \cdots
  \left[ \left( \mathcal{\tilde{T}}({\mbox{\boldmath$q$}_\alpha})\right)^{n_{\alpha-1}}\right]_{\eta_{\alpha-1}\eta_\alpha}
  \left(\mathcal{T}^{n_\alpha}\right)_{\eta_\alpha\eta_{\alpha+1}}\nonumber \\
&\ \ \ \times \delta_{0, N-m(\alpha+1)-(n_0+n_1+\cdots n_{\alpha+1})}.
\end{align}
\normalsize
By using {eqns.~(\ref{eqn:def_of_tail}) and (\ref{eqn:def_of_loop})}, we can derive eqn.~(\ref{eqn:conformation_single}).
\section{Behavior of $\langle M \rangle$ and driving force $\mu_{\rm c}$}\label{sec:M_and_mu}
In this appendix, we discuss the behavior of $\langle M \rangle$ which depends on $\mu_{\rm c}/T$, $\alpha$ and $m$.
We show the relationship between $\langle M\rangle$ and $\mu_{\rm c}/T$ in the cases {with} $(\alpha, m)=(0, 3), (1, 3)$ and $(2, 3)$, and {with} $\Delta\varepsilon/T=0.0$.
{As the value of $\alpha$ is relatively small,} the behavior of $\langle M \rangle$ does not depend on $\Delta \varepsilon / T$ in these sizes of the nuclei.
{Although $\langle M \rangle$ depends on $\Delta\varepsilon / T$ in the case of the large value of $\alpha$, this situation corresponds to the thick ordered region which is not consistent with the assumption of Model M.}
\begin{figure}[H]
  \begin{center}
    \includegraphics[width=10.0cm]{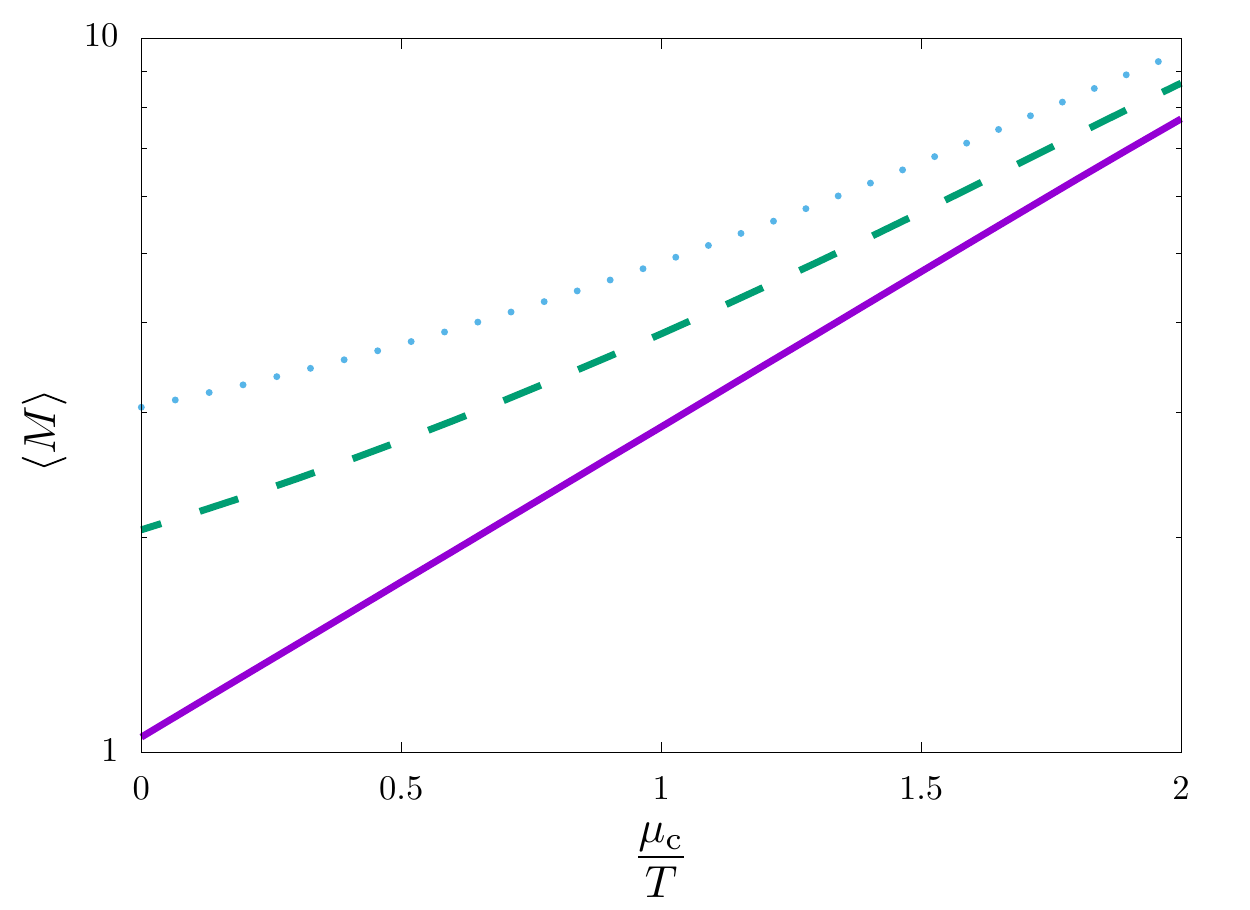}
    \caption{The dependence of the average number of chains $\langle M \rangle$ participating in the nucleus  on the chemical potential $\mu_{\rm c}/T$.
    {The vertical axis is logarithmic scale and the horizontal axis is linear scale.}
    The sizes of the nuclei {are} {$(\alpha, m)=(0, 3)$ (purple solid line), $(\alpha, m)=(1, 3)$ (green dashed line) and $(\alpha, m)=(2, 3)$ (blue dotted line)}.
    {The gauche energies are $\Delta\varepsilon/T=0.0$ for all sizes.}
    }
    \label{fig:number_of_average}
  \end{center}
\end{figure}
We notice the relationship between $\langle M \rangle$ and $\mu_{\rm c}/T$.
While $\langle M \rangle$ increases with $\mu_{\rm c}/T$ {, it} does not depend on the size of the nucleus.
$\langle M \rangle$ is roughly an exponential function of $\mu_{\rm c}/T$.
Thus, in the high $\mu_{\rm c}/T$ regime, the assumption of the small ordered region in the nucleus can not be justified.

We focus on the behavior of $\langle M \rangle$ at $\mu_{\rm c}/T=0$.
As we mentioned in subsection~\ref{sec:result_of_Model_M}, $\mu_{\rm c}/T$ is regarded as a driving force for the polymer chain to {migrate} from the bulk phase to the nucleus.
Therefore, $\mu_{\rm c}/T=0$ means that the driving force does not exist {except for the contribution from the CNT terms}.
In Fig.~\ref{fig:number_of_average}, {at $\mu_{\rm c}/T=0$ with $(\alpha, m)=(0, 3)$ ({the purple solid line}),} we {obtain} $\langle M \rangle\simeq 1$ {which is the minimum value for creating the nucleus}.
The behavior of $\langle M \rangle$ around $\mu_{\rm c}/T=0$ does not depend on $m$ but on $\alpha$.
In the case of $\mu_{\rm c}/T=0$ with $\alpha=0$ (see {the purple solid line} in Fig.~\ref{fig:number_of_average}), $\langle M \rangle \simeq 1$ which is the minimum value {for} $M$ for creating the nucleus.
{In the case of $\alpha=1$ and 2 specified by the green dashed line and the blue dotted line, respectively, in Fig.~\ref{fig:number_of_average}}, we obtain $\langle M \rangle> \alpha$ at $\mu_{\rm c}/T=0$.
This result is explained as follows;
When the total number of the loops is specified by $\alpha$ in Model M, {the single chain tends not to include multiple loops} due to the large loss of the conformation entropy.
{In other words, taking the conformation entropy loss into account, the nucleus with $\alpha$ loops tend to be composed of at least $\alpha$ chains where a chain may or may not include a single loop.}
{Therefore, the average number of chains is larger than $\alpha$ at $\mu_{\rm c}/T=0$.}
\bibliographystyle{achemso}
\bibliography{ref}

\end{document}